\documentclass[sigconf,authorversion,noacm]{acmart}
\AtBeginDocument{%
  \providecommand\BibTeX{{%
    \normalfont B\kern-0.5em{\scshape i\kern-0.25em b}\kern-0.8em\TeX}}}

\usepackage{amsmath,amsfonts}
\usepackage{algpseudocode}
\usepackage{algorithm}
\usepackage{graphicx}
\usepackage{textcomp}
\usepackage{xcolor}
\usepackage{array}
\usepackage{rotating}
\usepackage{hyperref}
\usepackage{comment}
\usepackage{xspace}
\usepackage{xcolor}
\usepackage{listings}
\usepackage{longtable}
\usepackage{blindtext}
\usepackage{lscape}
\usepackage{supertabular}

\usepackage{pifont}% http://ctan.org/pkg/pifont
\usepackage{babel,booktabs,multirow}
\usepackage{lscape}
\usepackage{wasysym}
\usepackage{tikz}
\usepackage{subcaption}
\usepackage{longtable}

\newcommand*{\ie}{i.e.,\@\xspace}
\newcommand*{\eg}{e.g.,\@\xspace}

\newcommand*{\etal}{et al.\@\xspace}

\newcommand*{\tool}{SERAPHINE\@\xspace}
\newcommand*{\HGF}{Hugging Face\@\xspace}

\newcommand*{\totPTM}{381,240\@\xspace}

\newcommand*{\freqPTM}{135,915\@\xspace}

\newcommand*{\totTag}{40\@\xspace}

\newcommand*{\freqTag}{19\@\xspace}

\lstdefinestyle{searchstringstyle}{
  basicstyle=\ttfamily\scriptsize,
breaklines=true,                 
  captionpos=b,                    
  numbers=none,                    
  numbersep=5pt,                  
  showspaces=false,                
  showstringspaces=false,
  showtabs=false,                  
  tabsize=2,
  frame=single
}

\newcommand\revision[1]{\textcolor{black}{#1}}

\newcommand{\rqfirst}{\textbf{RQ$_1$}: \textit{How effectively can state-of-the-art classifiers categorize pre-trained models based on the information provided in their model cards?}}

\newcommand{\rqsecond}{\textbf{RQ$_2$}: \textit{How can the suggested mapping contribute to assisting developers in the selection of an appropriate pre-trained model?}}

 \newcommand*\circled[1]{\tikz[baseline=(char.base)]{\color{black} 
		\node[shape=circle,draw=black,fill=white!0!white,inner sep=.3pt] (char) {{{\texttt\textbf #1}}};}}

\newcommand{\mybox}[4]{
	\begin{figure}[h]
		\centering
		\begin{tikzpicture}%[boxrule=0.86pt,left=0.3em, right=0.3em,top=0.1em, bottom=0.05em]
			\node[anchor=text,text width=\columnwidth-0.5cm, draw, rounded corners, line width=0.5pt, fill=#3, inner sep=1mm] (big) {\\#4};
			\node[draw, rounded corners, line width=.2pt, fill=#2, anchor=west, xshift=1mm] (small) at (big.north west) {#1};
		\end{tikzpicture}
	\end{figure}
}

\begin{document} \sloppy


\title{\tool: Shepherding developers in select, test, and deploy pre-trained models }

\title{A semi-automated approach to map pre-trained models to software engineering tasks}

%\title{Towards automatically categorizing pre-trained models in software engineering: A case study with a \HGF dataset}

\title{Automated categorization of pre-trained models in software engineering: A case study with a \HGF dataset}

\title{Automated categorization of pre-trained models for software engineering: A case study with a \HGF dataset}

%\title{Combining Hierarchical Feature Models and Code-generation (?) for Configurable Travel Recommender Systems}

%\title{ Providing semi-automatic tourism planning using feature models \CDS{tentative title} }

%\title{ Towards Configurable Tourism Recommender Systems Using Multi-level Modeling \CDS{tentative title} }


\author{Claudio Di Sipio}
\email{claudio.disipio@univaq.it}
%\orcid{1234-5678-901}

\affiliation{%
  \institution{University of L'Aquila}
 % \streetaddress{P.O. Box 1212}
	\city{L'Aquila}
	\country{Italy}
}

\author{Riccardo Rubei}
\email{riccardo.rubei@univaq.it}
%\orcid{1234-5678-901}

\affiliation{%
  \institution{University of L'Aquila}
% \streetaddress{P.O. Box 1212}
	\city{L'Aquila}
	\country{Italy}
}

\author{Juri Di Rocco}
\email{juri.dirocco@univaq.it}
%\orcid{1234-5678-901}

\affiliation{%
  \institution{University of L'Aquila}
% \streetaddress{P.O. Box 1212}
\city{L'Aquila}
\country{Italy}
}

\author{Davide Di Ruscio}
\email{davide.diruscio@univaq.it}
%\orcid{1234-5678-901}

\affiliation{%
  \institution{University of L'Aquila}
% \streetaddress{P.O. Box 1212}
\city{L'Aquila}
\country{Italy}
}

\author{Phuong T. Nguyen}
\email{phuong.nguyen@univaq.it}
%\orcid{1234-5678-901}

\affiliation{%
  \institution{University of L'Aquila}
% \streetaddress{P.O. Box 1212}
\city{L'Aquila}
\country{Italy}
}

\renewcommand{\shortauthors}{Di Sipio et al.}

\begin{abstract}
Software engineering (SE) activities have been revolutionized by the advent of pre-trained models (PTMs), defined as large machine learning (ML) models that can be fine-tuned to perform specific SE tasks. However, users with limited expertise %in this field 
may need help to select the appropriate model for their current task. To tackle the issue, the Hugging Face (HF) platform simplifies the use of PTMs by collecting, storing, and curating several models. Nevertheless, the platform currently lacks a comprehensive categorization of PTMs designed specifically for SE, %software engineering, 
\ie the existing tags are more suited to generic ML categories.

This paper introduces an %a preliminary 
approach to address this gap by enabling the automatic classification of PTMs for %the context of 
SE tasks. First, we utilize a public %publicly available 
dump of HF to extract PTMs %PTM-related 
information, including model documentation and associated tags. Then, we employ a semi-automated method to identify SE tasks and their corresponding PTMs from existing literature. The %proposed 
approach involves creating an initial mapping between HF tags and specific SE tasks, using a similarity-based strategy to identify PTMs with relevant tags. 
% We eventually propose an initial map the generic tag with the specific SE task based on similarity-based strategy to retrieve similar PTM with the corresponding tag. 
%to create a labeled dataset where the model card is linked with the actual task. Finally, we train a textual classifier to predict the SE task given the PTM description. 
The %preliminary 
evaluation shows that model cards are informative enough to classify PTMs considering the pipeline tag. Moreover, we provide a %an initial 
mapping between SE tasks and stored PTMs %on HF 
by relying ons model names.
%, though an in-depth analysis is needed to confirm the practicality of our approach. %can be used in practice. 

\end{abstract}
%\begin{CCSXML}
%<ccs2012>
%  <concept_id>00000000.0000000.0000000</concept_id>
%  <concept_desc>Do Not Use This Code, Generate the Correct Terms for %Your Paper</concept_desc>
%  <concept_significance>500</concept_significance>
%  <concept_id>00000000.00000000.00000000</concept_id>
%  <concept_desc>Do Not Use This Code, Generate the Correct Terms for %Your Paper</concept_desc>
%  <concept_significance>300</concept_significance>
%  <concept_id>00000000.00000000.00000000</concept_id>
%  <concept_desc>Do Not Use This Code, Generate the Correct Terms for %Your Paper</concept_desc>
%  <concept_significance>100</concept_significance>
%  <concept_id>00000000.00000000.00000000</concept_id>
%  <concept_desc>Do Not Use This Code, Generate the Correct Terms for %Your Paper</concept_desc>
%  <concept_significance>100</concept_significance>
%</ccs2012>
%\end{CCSXML}

%\ccsdesc[500]{Do Not Use This Code~Generate the Correct Terms for Your %Paper}
%\ccsdesc[300]{Do Not Use This Code~Generate the Correct Terms for Your %Paper}
%\ccsdesc{Do Not Use This Code~Generate the Correct Terms for Your Paper}
%\ccsdesc[100]{Do Not Use This Code~Generate the Correct Terms for Your %Paper}



\maketitle

\section{Introduction}
\label{sec:Introduction}
\begin{figure*}[t!]
	\centering
	\begin{subfigure}[b]{0.45\linewidth}
		\includegraphics[width=\linewidth]{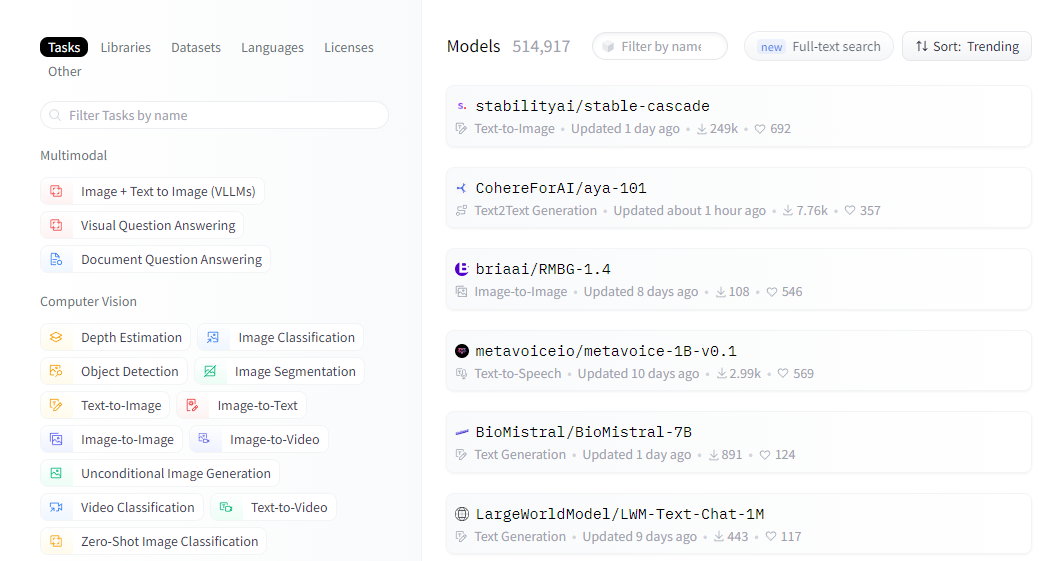}
		\caption{Selection mechanism based on pipeline tags}
		\label{fig:example}
	\end{subfigure}
	\hfill
	\begin{subfigure}[b]{0.40\linewidth}
		\includegraphics[width=\linewidth]{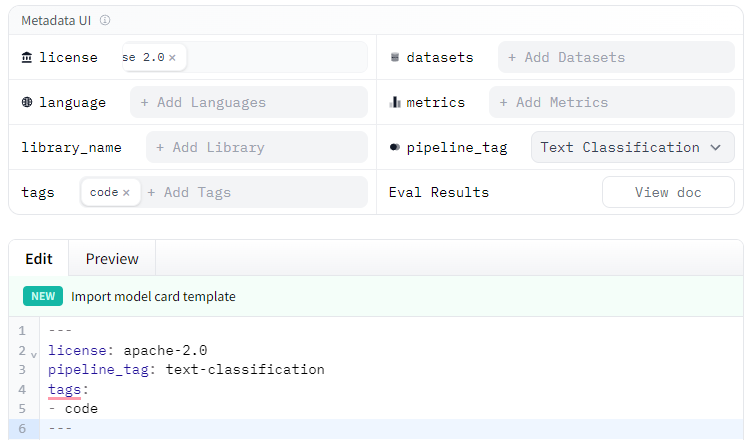}
		\caption{Model card configuration}
		\label{fig:modelcard}
	\end{subfigure}
	\caption{The \HGF PTM reuse-oriented capabilities.}
	\label{fig:combined}
\end{figure*}

The emergence of foundation models (FMs), \eg large language models (LLMs) \cite{hou_large_2023} or pre-trained models (PTMs) \cite{HAN2021225}
has significantly transformed traditional software engineering (SE) practices.  Interestingly, PTMs are playing a crucial momentum while being used as a promising technology to support a wide range of SE tasks, including domain analysis, source code development, or testing. 
%
%In software development,
%In recent years, we have witnessed a radical change in the traditional workflow that has been moved from manual and error-prone methodologies to automatic approaches based on foundation models (FMs), \eg large language models (LLMs) \cite{hou_large_2023,10109345} or pre-trained models (PTMs) \cite{HAN2021225}. The lattters have played a crucial role in this transition by providing more and more specialized strategies to implement SE tasks. 
%
Although recent studies have shown that PTMs can outperform traditional approaches in terms of accuracy \cite{tufano_using_2022,ding_can_2022,zhang_using_2022}, selecting the appropriate model for a given task remains a challenge for non-expert users. To alleviate the burden of choice, the developers can rely on \textit{model repositories} defined as dedicated open-source software repositories that store, collect, and maintain FMs \cite{gong_what_2023}. Furthermore, most of those community-based platforms offer dedicated capabilities to search and filter stored models, \eg tagging system or model documentation. Despite this, a proper mapping between the generic machine learning tasks and specific SE tasks is still missing.

In this paper, we present a semi-automated approach that centers around constructing a mapping between SE tasks and pre-trained models (PTMs) stored on the Hugging Face (HF) model repository,\footnote{\url{https://huggingface.co/}} recognized for housing the largest number of PTMs compared to other platforms \cite{gong_what_2023}. Additionally, the HF community project \cite{ait_hfcommunity_2023} represents an initiative that regularly updates the platform's dump, thereby facilitating empirical analyses of the stored PTMs. In particular, we aim at answering the following research questions:

%\begin{itemize}
%    \item 
\medskip
\noindent
\rqfirst 
	%\item 
	
\smallskip
\noindent	
\rqsecond 
\medskip
%\end{itemize}
 
%    \item \rqthird

%In this research question, we aim to construct a SE tasks taxonomy built on top of the existing work in the software engineering. In addition, we run the pseudo-algorithm describe in Section \ref{sec:mapping} to give an intuition on how the mapping can be realized in practice

To address $RQ_1$, we %our objective is to 
evaluate the informativeness of the model card accompanying each stored pre-trained model (PTM) in facilitating traditional classifiers. Specifically, we employ two state-of-the-art \revision{text multi-label} classifiers, namely the Complement Naive Bayesian (CNB) network \cite{rennie_tackling_nodate} and the C-Support Vector classifier (SVC) \cite{chang_libsvm_2011}. 
To answer \textbf{$RQ_2$}, we establish a taxonomy of SE %software engineering (SE) 
tasks, building upon existing work %in the field 
through a lightweight systematic study that considers the available literature. Utilizing the %collected 
data, we formulate an initial version of the mapping using a similarity-based algorithm. Our preliminary findings indicate that the provided model cards are sufficiently informative for automatic classification, relying on the associated tags. However, it is noteworthy that more than half of the PTMs lack corresponding cards, leaving a substantial number of models on \HGF uncovered. Furthermore, we present an initial iteration of the mapping between SE tasks and the generic PTM category. This mapping can be refined to encompass additional SE tasks or cater to different PTM repositories.

We suppose that devising a set of more advanced techniques to recommend the proper PTM given a specific SE task is crucial to support the creation of SE task-dedicated AI-based agents framework \cite{hong2023metagpt,dong2023selfcollaboration}. Thus, %In this respect, 
this paper can be seen as the very first stepping stone to enable dedicated recommender systems for software engineering (RSSEs) \cite{robillard_recommendation_2014,di_rocco_development_2021} that are able to automatically select the proper models given the SE task. 
It is our firm belief that devising advanced techniques for recommending the most suitable pre-trained model (PTM) based on a specific SE %software engineering (SE) 
task is imperative to sustain the establishment of frameworks consisting of AI-based agents that are tailored for specific SE tasks \cite{hong2023metagpt,dong2023selfcollaboration}. %In this context, 
Our paper serves as a %an initial 
foundational step, paving the way for the creation of dedicated RSSEs %recommender systems for software engineering (RSSEs) 
\cite{robillard_recommendation_2014,di_rocco_development_2021} endowing the capability to autonomously identify and recommend appropriate models tailored to given SE tasks. The main contributions of this paper are as follows:

\noindent \ding{228} An initial mapping between generic PTM tags and SE tasks by adopting a semi-automatic screening of existing literature;

\noindent \ding{228} An empirical evaluation conducted using the textual documentation available on \HGF to train state-of-the-art classifiers;

\noindent \ding{228} A publicly available replication package to foster further studies in this domain \cite{replicationPackage}.

%The paper is structured as follows. Section \ref{sec:Motivation} presents the typical workflow for using PTMs and the features of the HF platform. In Section \ref{sec:Methodology} we present our approach to mapping SE tasks to PTM-specific categories. The preliminary results of our study are presented in Section \ref{sec:evaluation} and possible threats to validity are discussed in Section \ref{sec:Threats}. We discuss the related work in Section \ref{sec:related} and conclude the paper in Section \ref{sec:Conclusion}. 

\section{Motivation and Background}
\label{sec:Motivation}

\subsection{Overview of the PTM reuse workflow}

As part of the \textit{decision-making PTM reuse} process outlined by Jiang et al. \cite{jiang_empirical_2023}, developers undertake distinct steps to search for and identify PTMs %pre-trained models 
suitable for supporting specific tasks. Initially, developers engage in a \textit{reusability assessment}, requiring a comprehensive analysis of the requirements to select the most appropriate PTM. % from the examined repository.
 Then, during the \textit{model selection} phase, developers %aim to 
identify a set of candidate PTMs. Following this, developers may evaluate whether the chosen PTM is suitable for deployment through the \textit{downstream evaluation}, which typically involves a fine-tuning phase specific to the task at hand. Finally, the \textit{model deployment} phase encompasses testing the finalized version of the PTM in a real development environment.
%\noindent
%\textbf{Downstream evaluation:} This phase aims to assess if the PTM is ready for the actual deployment. Several attempts may be needed to find the proper model or configuration, \ie the chosen PTMs need to be adapted most of the time using fine-tuning.

%\noindent
%\textbf{Model deployment:}  The last step is the deployment of the PTM in a real environment. The computation resources available play a crucial role since the selected PTM can be demanding in terms of time or efficiency. 

Even though this process is facilitated by model repositories and other open-source platforms such as GitHub, the presented workflow still considers generic categories (\eg object detection, image to text, text generation, and image classification) that are not directly linked to SE tasks. In this respect, we propose a taxonomy as a starting point to improve the first two phases of the above mentioned process, \ie \emph{reusability assessment} and \emph{model Selection}, for SE tasks. 
%An initial attempt to create a SE categorization of tasks has been done in \cite{gong_what_2023}. 

\subsection{The \HGF model repository}

To allow for their practical usage in SE tasks, PTMs need to be stored, maintained, and documented. In this respect,  developers can share their PTMs on model repositories. Among the others, \HGF offers the largest number of PTMs and the corresponding documentation. Furthermore, the \emph{HFCommunity} project \cite{ait_hfcommunity_2023} allows the analysis of metadata and source code contained in \HGF. 

Figure \ref{fig:combined} shows the model searching capabilities provided by \HGF. In particular, an interested developer can browse the hub by directly using the search bar. However, this process is time-consuming given the large number of PTMs stored on the platform. To mitigate this, HF provides a tagging system that can automatically filter the models by category as depicted in Figure \ref{fig:example}. For instance, the user can obtain the list of models that perform \texttt{text-classification}.

To further improve the visibility, PTM owners can upload the relevant information in the \textit{model card}, \ie a README-like document that provides the information required to configure and run the PTM at hand \cite{10.1145/3287560.3287596}. Figure \ref{fig:modelcard} represents the creation process of a model card using the dedicated interface. 
In addition, the user who wants to publish a model can select two types of tags, \ie \texttt{pipeline tag} and \texttt{user-based tags}. The former represents the community-based tagging system that expresses generic tasks, \eg image classification, text-generation. Users manually specify the latter, like the GitHub topics mechanism.\footnote{\url{https://github.com/topics}} 

 %Furthermore, this type of tag identifies the primary goal of the PTMs. %The extraction and analysis of the user-based tags can be seen as a possible future work.  
Unfortunately, many models stored in HF are not documented, or present missing information and discrepancies \cite{jiang_empirical_2023,castano_analyzing_2023}. Moreover, non-expert developers may need help to elicit the proper PTM according to the SE task of interest. Thus, we see an urgent need to provide developers with a rigorous mapping between HF tags and SE-related activities. In the scope of this paper, we focus on mapping pipeline tags to SE tasks since they are maintained by HF, thus avoiding erroneous labeling by users.

%Even though the \HF platform offers several capabilities to filter the PTMs, the corresponding information is uploaded by the developers, generating in some cases discrepancies \cite{montes_discrepancies_2022}. 

\section{Proposed approach}
\label{sec:Methodology}

%\begin{figure}[h!]
%    \includegraphics[width=\linewidth]{figures/architecture.pdf}
%    \caption{The architecture of the envisioned approach}
%    \label{fig:approach}
%\end{figure}

%Given the task under development, the framework automatically retrieves the first set of candidate models based on the textual description. To this end, we exploit a classifier module that can be trained on HF data, thus retrieving the most similar model compared to the description provided by the developer. Once the task has been identified, the approach searches for additional data related to that model, \ie input data, hardware prerequisites, and related discussion available on HF. The chosen model is eventually tested using the available data to collect quality measurements, \ie energy consumption, accuracy, and training time. 

%In this section, we present the SERAPHINE framework aims to support the developers in selecting and deploying PTMs according to a set of user requirements, e.g., the size of the model, the computation time, or the energy consumption. 

\begin{figure}[h!]
	\includegraphics[width=0.88\linewidth]{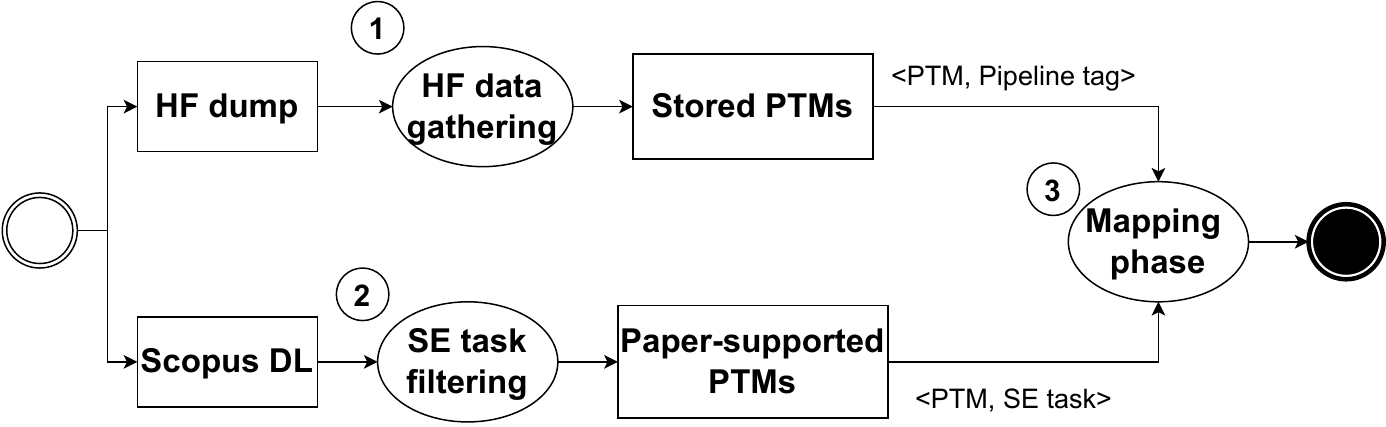}
	\caption{The proposed mapping approach.}
	\vspace{-.2cm}
	\label{fig:approach}
\end{figure}

Figure \ref{fig:approach} shows the proposed mapping approach: the data filtering phase is split into two main phases, \ie  \emph{HF data gathering} \circled{1}  and \emph{SE task filtering} \circled{2}. A \emph{mapping phase} \circled{3} is subsequently performed to correlate identified SE tasks with candidate pre-trained models stored on Hugging Face. The subsequent sections elaborate on these three core phases.

\subsection{\textbf{HF data gathering}} \label{sec:hf_data}

To gather the necessary data, we utilized the latest available dump from the HF community website,\footnote{\url{https://som-research.github.io/HFCommunity/download.html}} specifically from October 2023. We deployed this dump locally into a MySQL database to expedite the data retrieval process. Within the scope of this paper, our focus is on extracting textual documentation, as represented by the model card, and the \textit{pipeline tags}. Additionally, we collected information for each pre-trained model, including the number of likes and downloads, facilitating further qualitative analysis discussed in Section \ref{sec:evaluation}. For these purposes, we employed the dedicated Python library\footnote{\url{https://pypi.org/project/mysql-connector-python/}} to interact with the Hugging Face dump.

\begin{lstlisting}[style=searchstringstyle,firstnumber=1,caption=SQL query., captionpos=t, label={lst:sql},numbers=none]
  SELECT model_id,card_data,pipeline_tag,likes,downloads FROM model,repository WHERE model.model_id = repository.id;
  \end{lstlisting}
  
The SQL query used for this interaction is provided in Listing \ref{lst:sql}. This process resulted in a CSV file containing 381,240 PTMs along with the aforementioned metadata.

%In addition, we applied different quality filters to support the mapping with the SE tasks. The details of such preprocessing are discussed in Section \ref{sec:evaluation}

\subsection{\textbf{SE task filtering}} \label{sec:se_tasks}

\begin{figure}[h!]
	\includegraphics[width=0.88\linewidth]{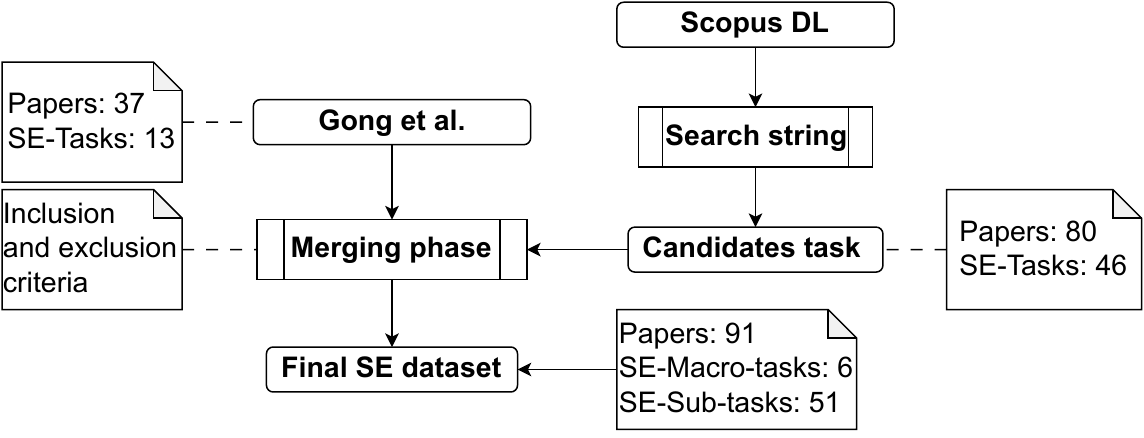}
	\vspace{-.3cm}
	\caption{The performed process to elicit SE tasks.}
	\vspace{-.2cm}
	\label{fig:mapping}
\end{figure}

We followed the steps shown in Figure \ref{fig:mapping} to collect SE tasks that are supported by PTMs. In particular, the process starts from the initial work by Gong \etal \cite{gong_what_2023} 
identifying 13 SE tasks supported by PTMs from 37 papers. Though it is recent work, we expanded this initial mapping since PTM-related technologies are fast-evolving \cite{10.1145/3453478}. To this end, we executed a query on the Scopus digital library\footnote{\url{https://www.scopus.com/}} with the following set of keywords: \textit{(i)} pre-trained model* OR PTM* OR large language model* OR LLM OR transformer*; \textit{(ii)} AUT support* OR  recommend* OR task* OR automat*; \textit{(iii)} requirement* OR develop* OR source code. All such keywords are combined by using the AND operator. We considered papers published over the last five years in top-tier SE venues, \eg ICSE, ASE, and TSE. Through the executed query, we retrieved 80 papers and 46 tasks for which PTMs were employed. Afterwards, we refined the result by \textit{(i)} merging the papers and SE tasks that were in common with those defined by Gong \etal, and \textit{(ii)} applying a set of inclusion and exclusion criteria to the titles and abstracts. In particular, we included all the approaches that reuse PTMs to cover different SE tasks, \eg code completion, bug fixing, or vulnerability assessment.  
In contrast, foundational papers on PTMs and works that combine PTMs with other techniques were filtered out from the final list. In addition, we excluded empirical studies that investigate qualitative aspects of PTMs, \eg carbon emission or advanced fine-tuning strategies.

We eventually identified 91 papers and 51 SE tasks. Since some of the collected papers address a very specific task, \eg \textit{Smart contract generation} or \textit{Software traceability}, we refined those categories by \revision{manually grouping SE tasks into 6 macro tasks}\footnote{Interested readers can refer to the complete list of macro and sub-tasks in the online appendix \url{https://tinyurl.com/2p8y25cr}} as reported in Section \ref{sec:evaluation}, \eg \textit{Code-related tasks} or \textit{Documentation support}.

%Due to the lack of space, the interested reader can find the complete list of macro and sub-tasks in the online appendix publicly available \footnote{\url{here}}. 

%only 140,149 do not have null entries in the two relevant fields for our study, \ie the pipeline tag and model card. 

\subsection{\textbf{Mapping phase}} \label{sec:mapping}

%In this section, we give the intuition of the mapping strategy. Performing a comprehensive mapping between the identified SE task and every single PTM stored on \HGF is out of the scope of this paper. 

The proposed mapping algorithm is represented in Algorithm \ref{alg:cap}. While a comprehensive mapping between each identified SE task and every individual pre-trained model stored on \HGF is out of the scope of this paper, we present a generic approach that holds applicability to model repositories resembling HF.

%\begin{algorithm}
%  \caption{Match and Assign Pipeline Tags}
%  \begin{algorithmic}[1]
%  \For{each pair $\langle PTM, SE\, task \rangle$}
%      \If{match$(PTM, PTM^*)$}
%          \State $PipelineTag \gets SE\, task$
%      \EndIf
%  \EndFor
%  \Function{match}{$PTM, PTM^*$}
%      \State $\alpha \gets sim(PTM, PTM^*)$
%      \If{$\alpha > threshold$}
%          \State \Return true
%      \Else
%          \State \Return false
%      \EndIf
%  \EndFunction
%  \end{algorithmic}
%  \end{algorithm}
%  

%\begin{algorithm}
%  \caption{Pseudocode for the mapping phase}\label{alg:cap}
%  \begin{algorithmic}
%    \Procedure{MapPipelineTag}{$PTM, SE\,task, ListOfPTM^*$}
%        \For{$\langle PTM^*, PipelineTag \rangle$ in $ListOfPTM^*$}
%            \If{\Call{match}{$PTM, PTM^*$}}
%                \State $PipelineTag \gets SE\, task$
%            \EndIf
%        \EndFor
%    \EndProcedure
%    \Function{match}{$PTM, PTM^*$}
%        \State $\alpha \gets$ \Call{sim}{$PTM, PTM^*$}
%        \If{$\alpha > threshold$}
%            \State \Return true
%        \Else
%            \State \Return false
%        \EndIf
%    \EndFunction
%  \end{algorithmic}
%\end{algorithm}

\begin{algorithm}
  \caption{Pseudocode for the mapping phase}\label{alg:cap}
  \footnotesize
  \begin{algorithmic}[1]
    \State \textbf{Input:} $\langle ptm:PTM, t:SETask, HF \rangle$ 
    \State \textbf{Output:} $MAPPING$
    \State $MAPPING  \gets \{ \}$    
    \For{$ptm_i \in HF$}
        \If{\Call{match}{$ptm_i, ptm$}}
            \State $MAPPING.add\left(\langle ptm_i.pipelinetag, t \rangle\right)$ 
        \EndIf
    \EndFor
    \Function{match}{$PTM, PTM^*$} \label{lst:fun:match}
        \State $\alpha \gets$ \Call{sim}{$PTM, PTM^*$}
        \State \Return $\alpha > T$
    \EndFunction
  \end{algorithmic}
\end{algorithm}

%\begin{algorithm}
%  \caption{Pseudocode for the mapping phase}\label{alg:cap}
%  \footnotesize
%  \begin{algorithmic}[1]
%    \State \textbf{Input:} $\langle ptm:PTM, t:SETask, HF \rangle$ 
%    \State \textbf{Output:} $MAPPING$
%    \State $MAPPING  \gets \{ \}$    
%    \For{$ptm_i \in HF$}
%        \If{\Call{match}{$ptm_i, ptm$}}
%            \State $MAPPING.add\left(\langle ptm_i.pipelinetag, t \rangle\right)$ 
%            \EndIf
%    \EndFor
%    \Function{match}{$PTM, PTM^*$} \label{lst:fun:match}
%        \State $\alpha \gets$ \Call{sim}{$PTM, PTM^*$}
%        \If{$\alpha > T$}
%            \State \Return true \Else
%            
%            \State \Return false
%        \EndIf
%    \EndFunction
%  \end{algorithmic}
%\end{algorithm}

%The algorithm takes as input a pre-trained model $ptm$, the corresponding SE task $t$ that has been found supported by $ptm$ according to the study discussed in the previous section, and the whole $HF$ dump (see line 1). For each PTM stored in the considered repository, its name is 
%analyzed and the similarity with the name of the input  \texttt{ptm} is calculated. If their similarity is greater than the threshold \texttt{T} then it means that $ptm_i$ is a pre-trained model that should be used to support the SE task $t$ in input. Consequently, the mapping between the pipeline tag of $ptm_i$ and $t$ is established. 

As shown in Algorithm \ref{alg:cap}, there are three main inputs: a pre-trained model denoted as $ptm$, a corresponding SE task labeled as $t$, and the entire repository $HF$ (Line 1). For each pre-trained model $ptm_i$ stored in the specified repository (Line 4), the algorithm analyzes its name and calculates the similarity with the name of the input \texttt{ptm} (Line 5). If the similarity surpasses the designated threshold \texttt{T} (Line 11), it signifies that $ptm_i$ is a relevant pre-trained model to support the SE task $t$. Consequently, the algorithm establishes a mapping between the pipeline tag of $ptm_i$ and the SE task $t$ provided as input (Line 6).

Intuitively, given the PTM name that is applied on a given SE task, we try to find a list of stored PTMs using the similarity between the given model name and different versions available on the dump. In this paper, we employed the Levenshtein distance metric \cite{levenshtein} as similarity function with $T$=0.8 as threshold.

  \begin{lstlisting}[style=searchstringstyle,firstnumber=1,caption=Example of the similar models retrieved for \texttt{RoBERTa}, captionpos=t, label={lst:mapping},numbers=none]
('sloberta', 'fill-mask'),
('roberta-go', 'fill-mask')
('me-roberta', 'fill-mask')
('am-roberta', 'fill-mask')
('numroberta', 'fill-mask')
    \end{lstlisting}

    Listing \ref{lst:mapping} shows an explanatory result, which is obtained by considering RoBERTa \cite{liu2019roberta} as ptm model and  Code related-task as SE task. By running the query on the HF dump, we got that the similar PTMs are labeled as \texttt{fill-mask}. Thus, we can establish a mapping between the pipeline tag \texttt{fill-mask} with the \texttt{Code-related SE} task.

It is worth noting that the PTMs belonging to the same family can have different pipeline tags. Similarly, the same model can be employed for different SE tasks. We try to mitigate these issues by \textit{(i)} filtering HF data by using quality filters and \textit{(ii)} grouping similar SE tasks into macro tasks. The results of the mapping are discussed in Section \ref{sec:evaluation}.

\section{Preliminary evaluation}
\label{sec:evaluation}
%\subsection*{\rqfirst} \label{sec:rq1}
\paragraph*{\textbf{Addressing $RQ_1$}}

%To answer this research question, we first filtered the initial dump discussed in Section \ref{sec:hf_data} by applying widely adopted preprocessing steps in the context of automatic categorization to increase the overall accuracy \cite{di_rocco_hybridrec_2023,izadi_semantically-enhanced_2023}.

%Table \ref{tab:filtering} summarizes the filtering steps to obtain the final dataset, \ie $D_f$. First, we removed PTMs that were missing relevant data, \ie pipeline tags or the model card. Afterward, we removed an additional set of PTMs and tags by considering two different thresholds, \ie support=$\alpha$ and downloads=$\beta$ where $\alpha$ is the median value and $\beta$ the mean value of the corresponding parameters. In such a way, we reduced the number of tags to enable more accurate predictions. 

To address this %research 
question, we initiated the process by refining the initial dump %presented 
in Section \ref{sec:hf_data}, through the application of well-established preprocessing steps commonly utilized in the realm of automatic categorization for heightened overall accuracy \cite{di_rocco_hybridrec_2023,izadi_semantically-enhanced_2023}.

\begin{table}[t!]
	\centering
	\footnotesize
	\caption{Filtering process to create $D_f$. }
	\vspace{-.4cm}
	\begin{tabular}{|l| c | c |}
		\hline
		& \textbf{\#PTMs} & \textbf{\#pipeline tags}  \\ \hline
		\textbf{PTMs in the initial dump}& \totPTM & \totTag
		\\ \hline 
		{\textbf{PTMs with missing data}} & 241,091 & - \\  \hline    
		
		{\textbf{PTMs with support$\le\alpha$ and downloads$\le\beta$}} & 4,234 & 21 \\ \hline 
		
		\textbf{$D_f$} & \textbf{\freqPTM} & \textbf{\freqTag}
		\\ \hline
	\end{tabular}
	\label{tab:filtering}
\end{table}

%<<<<<<< HEAD
%The steps involved in filtering and shaping the final dataset, denoted as $D_f$, are outlined in Table \ref{tab:filtering}. Initially, we excluded PTMs lacking essential data, such as pipeline tags or the model card. Subsequently, a further set of PTMs and tags were eliminated, employing two distinct thresholds: support=$\alpha$ and downloads=$\beta$, with $\alpha$ representing the median value and $\beta$ signifying the mean value of the corresponding parameters (the attribute ``support'' refers to the number of models for a specific category, while ``downloads'' pertains to the number of downloads for a given PTM). \textbf{DAVIDE: The thresholds are considered in AND or OR?}
%=======
The steps involve filtering and shaping the final dataset, denoted as $D_f$ (see Table \ref{tab:filtering}). Initially, we excluded PTMs lacking essential data, such as pipeline tags or the model card. Subsequently, a further set of PTMs and tags were eliminated, employing two distinct thresholds: support=$\alpha$ and downloads=$\beta$, \revision{where $\alpha$ is the median value and $\beta$ is the mean value of the corresponding parameters (the attributes ``support'' and ``downloads'' refer to the number of models for a specific category, and %while ``downloads'' pertains to 
the number of downloads for a given PTM, respectively)}.
%>>>>>>> b311ee956a8a3b45e4eac13f79133005ef2ff1f1

%``${public\ synchronized\ string}$''

\begin{table}[h!]
	\centering
	\footnotesize
	\caption{D$_f$ Results.}
		\vspace{-.2cm}
	\label{tab:Df_results}
	\begin{tabular}{@{}c*{6}{c}@{}}
		\toprule
		& \multicolumn{2}{c}{\textbf{Precision}} & \multicolumn{2}{c}{\textbf{Recall}} & \multicolumn{2}{c}{\textbf{F1 Score}} \\
		\cmidrule(lr){2-3} \cmidrule(lr){4-5} \cmidrule(lr){6-7}
		\textbf{Fold} & \textbf{CNB} & \textbf{SVC} & \textbf{CNB} & \textbf{SVC} & \textbf{CNB} & \textbf{SVC} \\ \midrule
		1 & 0.892 & \textbf{0.940} & 0.890 & \textbf{0.938} & 0.887 & \textbf{0.938} \\
		2 & 0.888 & \textbf{0.935} & 0.886 & \textbf{0.933} & 0.882 & \textbf{0.932} \\
		3 & 0.893 & \textbf{0.936} & 0.889 & \textbf{0.933} & 0.886 & \textbf{0.933} \\
		4 & 0.891 & \textbf{0.939} & 0.889 & \textbf{0.936} & 0.886 & \textbf{0.936} \\
		5 & 0.892 & \textbf{0.936} & 0.889 & \textbf{0.933} & 0.886 & \textbf{0.933} \\
		6 & 0.892 & \textbf{0.936} & 0.891 & \textbf{0.934} & 0.888 & \textbf{0.934} \\
		7 & 0.887 & \textbf{0.934} & 0.883 & \textbf{0.931} & 0.880 & \textbf{0.931} \\
		8 & 0.894 & \textbf{0.942} & 0.891 & \textbf{0.940} & 0.887 & \textbf{0.940} \\
		9 & 0.890 & \textbf{0.939} & 0.887 & \textbf{0.936} & 0.883 & \textbf{0.936} \\
		10 & 0.887 & \textbf{0.938} & 0.884 & \textbf{0.935} & 0.881 & \textbf{0.935} \\
		\midrule
		\textbf{Average} & 0.891 & \textbf{0.938} & 0.888 & \textbf{0.935} & 0.885 & \textbf{0.935} \\ \bottomrule
		
	\end{tabular}
\end{table}

%\subsection{Dataset and configuration settings}  \label{sec:settings}

%\begin{figure}
%    \includegraphics[width=1\linewidth]{figures/d1_distribution.pdf}
%    \caption{Distribution of tags in D1}
%    \label{fig:d1_distrbution}
%\end{figure}

%\begin{figure}
%    \includegraphics[width=1\linewidth]{figures/d2_new_distribution.pdf}
%    \caption{Distribution of tags in D2}
%    \label{fig:d2_distrbution}
%\end{figure}

%\begin{figure}
%    \includegraphics[width=1\linewidth]{figures/d2_sub_distribution.pdf}
%    \caption{Distribution of tags in D2\_sub}
%    \label{fig:d2_sub_distrbution}
%\end{figure}

Afterward, we performed the cross-fold validation \cite{Refaeilzadeh2009} of two state-of-the-art binary classifiers, \ie CNB and SVC models, and widely-adopted accuracy metrics, \ie \textit{Precision}, \textit{Recall}, and \textit{F1-score} \cite{buckland_relationship_1994}. Table \ref{tab:Df_results} shows the outcomes of the performed evaluation by considering the PTM stored on HF as the input and the \texttt{pipeline tag} as the predict label.

It is worth noting that the two employed models obtain decent performance, with the SVC model outperforming the CNB network by 10\% on average in terms of the considered accuracy metrics. 
However, we report that 241,091 PTMs in the dump have no model cards as shown in Table \ref{tab:filtering}. %This is an underpinning limit of the HF platform since the model card is not mandatory.  
%although it improves the reproducibility of PTMs \cite{10.1145/3287560.3287596}. 

\mybox{\textbf{\small{Answer to RQ$_1$}}}{gray!10}{gray!10}{\small{Our findings confirm that model cards can be used to enable traditional classifiers based on textual content, even though most of the developers do not upload this information on the platform.}}

%\begin{shadedbox}
%    \textbf{Answer to $RQ_1$:} Our findings confirm that model cards can be used to enable traditional classifiers based on textual content, even though most of the developers do not upload this information on the platform. 
%\end{shadedbox}

\smallskip
%\subsection*{\rqsecond} \label{sec:rq2}
\paragraph*{\textbf{Addressing $RQ_2$}}

%\begin{table*}[h]
%    \centering
%    \footnotesize
%    \begin{tabular}{|l| c | p{10cm}|}
%    \hline
%    \textbf{Macro-task} & \textbf{Support} & \textbf{Paper IDs} \\
%    \hline
%    Code-related tasks & 33 & 129, 2, 4, 7, 9, 10, 13, 14, 18, 20, 23, 24, 30, 34, 39, 41, 48, 50, 51, 53, 54, 55, 57, 60, 62, 65, 67, 76, 77, 106, 109, 123, 126 \\
%    \hline
%    Program repair & 16 & 1, 35, 36, 7, 40, 45, 110, 111, 113, 19, 20, 115, 24, 126, 27, 30 \\
%    \hline
%    Documentation support  & 22 & 132, 5, 7, 16, 17, 21, 22, 25, 29, 31, 33, 37, 49, 58, 59, 70, 74, 78, 105, 112, 119, 121 \\
%    \hline
%    Classification of SE artifacts & 8 & 45, 44, 49, 68, 117, 118, 12, 6 \\
%    \hline
%    Text-engineering tasks & 8 & 8, 32, 72, 47, 114, 43, 56, 71\\
%    \hline
%    Miscellaneous & 11 & 128, 3, 73, 42, 75, 107, 108, 26, 124, 125, 127 \\
%    \hline
%    \end{tabular}
%    \caption{Support and Paper IDs for Each Macro-task}
%    \label{tab:mapping_table}
%    \end{table*}
%

\begin{figure}[t!]
    \includegraphics[width=0.80\linewidth]{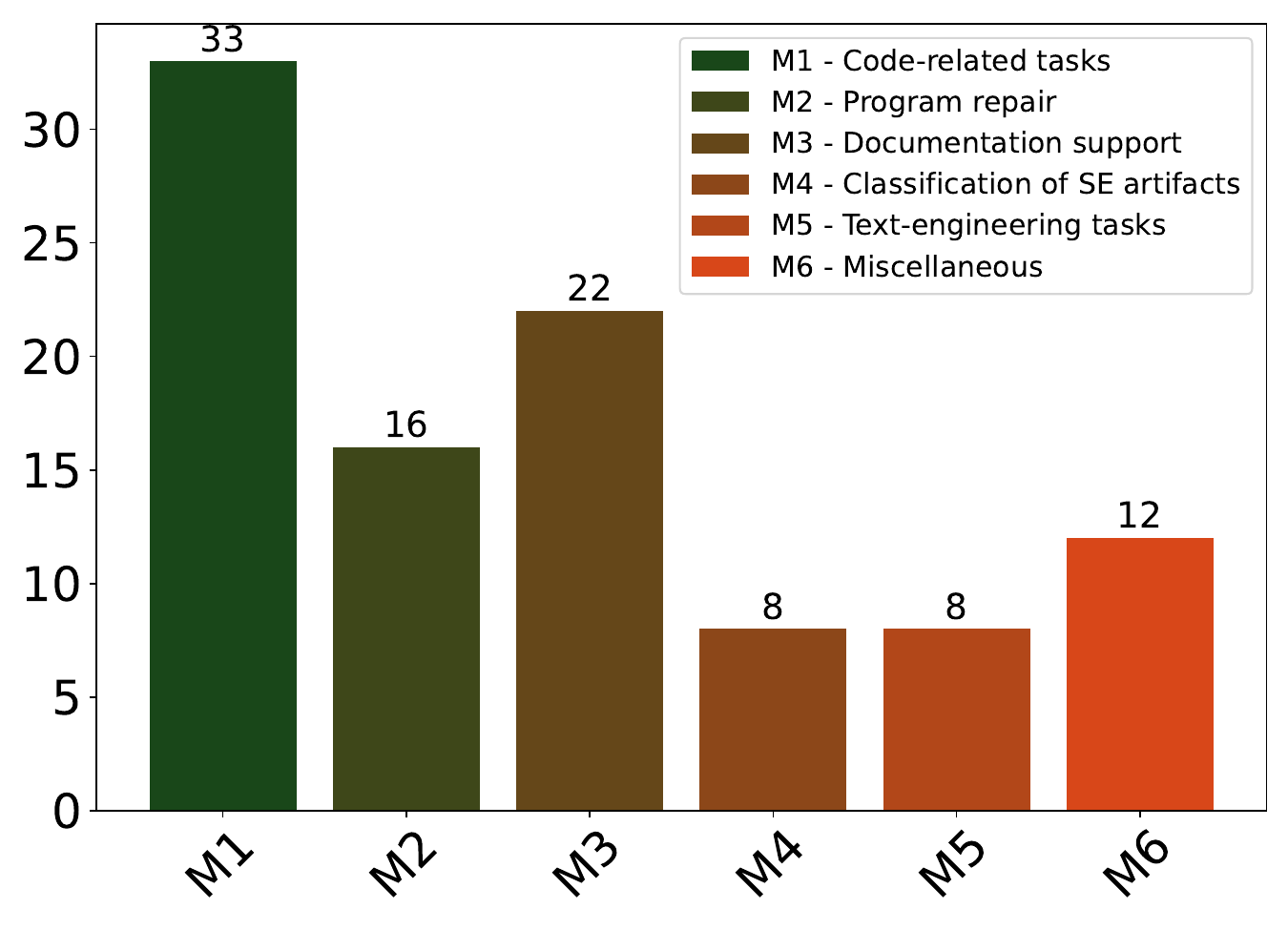}
    \vspace{-.2cm}
    \caption{The identified SE macro tasks.}
    \vspace{-.3cm}
    \label{fig:macro_stats}
\end{figure}

The outcome of the SE task filtering process described in Section \ref{sec:se_tasks} is reported in Figure \ref{fig:macro_stats}. In particular, we grouped similar SE tasks into six different \textit{macro-tasks} as follows:

\noindent \ding{228}~\textit{M1-Code-related tasks:} This category includes PTMs that have been used for several code-related tasks, from suggesting libraries to code summarization. %We include in this category also PTM which performs the code summarization and code search.

\noindent \ding{228}~\textit{M2-Program repair:} Compared to the previous category, we consider in this macro-task PTMs that have been applied to specific testing and program repair tasks \eg bug fixing or vulnerability detection in the code.
% Even though this task involves the source code as an SE artifact, most of the used PTMs have been fine-tuned to handle this specific class of SE activities

\noindent \ding{228}~\textit{M3-Documentation support:} We classify in this category PTMs that help developers generate textual documentation for different SE artifacts, \ie code, commits or bug reports. 

\noindent \ding{228}~\textit{M4-Classification of SE artifacts:} This category identifies PTMs that perform automatic classification of specific artifacts, \eg SO posts, commits, and issue reports.

\noindent \ding{228}~\textit{M5-Text-engineering tasks:} Unlike documentation-related tasks, we labeled PTMs that generate a textual description of specific SE artifacts, \eg titles for GitHub issues, Stack Overflow posts, or pull request descriptions. 

\noindent \ding{228}~\textit{M6-Miscellaneous:} This category includes PTMs that cannot be categorized in one of the previous macro-tasks. For instance, PTMs that support sentiment analysis in SE fall in this category. 

As expected, most of the PTMs are used to address \textit{code-related} tasks, \ie 33 papers employed them for code generation, code summarization, or code2code translation. On the contrary, the application to more generic tasks such as \textit{text-engineering} or \textit{classification of SE artifacts} is limited (we collected only 8 supporting papers for these two categories). Afterward, we run the algorithm shown in Section \ref{sec:mapping} for three well-adopted PTMs, \ie \texttt{BERT}, \texttt{RoBERTa}, and \texttt{T5}. To this end, we first retrieved a list of similar PTMs stored on HF to find the most frequent \texttt{pipeline tags} used on the dump. We eventually searched the PTM name in the abstract of the retrieved papers to get the corresponding SE macro-tasks and sub-tasks. The results are reported in Table \ref{tab:exampleMapping}.

%\begin{table}[h!]
%    \footnotesize
%    \centering
%    \begin{tabular}{|c| c| c c|}
%        \hline
%        \textbf{PTM} & \textbf{Pipeline tag} & \textbf{Macro SE task} & \textbf{Sub SE tasks} \\ \hline
%        \multirow{3}{*}{RoBERTa} & \multirow{3}{*}{fill-mask} & Code-related task & Code generation/completion, Algorithm classification, Code clone detection \\ 
%        & & Documentation/Requirements & Code search, API reviews classification \\
%        & & Miscellaneous, Classification of SE artifacts & Sentiment analysis, Issue report classification \\ \hline
%        T5 & & &\\ \hline
%        BART & & & \\ \hline
%    \end{tabular}
%    \caption{Explanatory mapping using our approach}
%    \label{tab:exampleMapping}
%\end{table}

\begin{table}[b!]
    \footnotesize
    \centering
         \vspace{-.2cm}
     \caption{Explanatory mapping using our approach.}
     \vspace{-.2cm}
    \begin{tabular}{|l| l| l | l|}
        \hline
        \textbf{PTM} & \textbf{Pipeline tag} & \textbf{Macro SE tasks} \\ \hline
        BERT & text-classification & M1,M2,M3,M4,M5,M6 \\ \hline
       RoBERTa & fill-mask & M1, M3, M4  \\ \hline
        T5 & text2text-generation & M1, M2, M3, M5 \\ \hline
    \end{tabular}
    \label{tab:exampleMapping}
\end{table}

%It is worth noting that 
Notably, \texttt{BERT} \revision{has been used to cover  
the whole set of SE macro tasks even though it may be not suitable for generation tasks}. Remarkably, we observe that \texttt{T5} is employed to address both code-related and documentation tasks, indicating the adaptability of text-to-text generation strategies even in the context of source code. The  trend reveals a significant surge in the utilization of PTMs across diverse SE tasks, extending beyond the realm of code-related activities. In this context, our approach is envisioned as an initial solution to aid developers in selecting a specific PTM tailored to the current SE task.

%\vspace{.1cm}
%\noindent
%$\rhd$~

%\begin{shadedbox}
%    \textbf{Answer to $RQ_2$:} 
%    Although we tested only three PTMs, the proposed mapping could be adopted to assist developers in selecting specific models that are used in practice. 
%\end{shadedbox}

\mybox{\textbf{\small{Answer to RQ$_2$}}}{gray!10}{gray!10}{\small{Although we tested only three PTMs, the proposed mapping could be adopted to assist developers in selecting specific models that are used in practice.}}

%\vspace{.1cm}
%\noindent
%$\rhd$~
% \textbf{Precision} is the ratio of number of matched elements to the number of recommended items: %for a metamodel:
%\begin{equation}
%precision = \frac{match(i)}{N}
%\end{equation}
%
%%\vspace{.1cm}
%%\noindent
%%$\rhd$~
%\textbf{Recall} is the ratio of the ground-truth items being found in the \textit{top-N} recommended ones:
%\begin{equation}
%recall = \frac{match(i)}{GT(i)} 
%\end{equation}
%
%
%%\vspace{.1cm}
%%\noindent
%%$\rhd$~
%\textbf{F-measure} is computed as the harmonic average of precision and recall:% by means of the following formula:
%\begin{equation}
%F-measure = \frac{2 * precison * recall }{precison + recall}
%\end{equation}

%subsection{Evaluation process}

%\subsection{Automatic classification}  \label{sec:classifiers}

%\section{Results}
%\label{sec:results}
%\input{src/Results}

\section{Threats to validity}
\label{sec:Threats}
%This section discusses the threats that could impact the validity of the obtained results and outlines the adopted mitigation strategies.
%
\textit{Internal validity} concerns the employed dump to perform the automatic classification given the model card, wherein the data may be incomplete or unbalanced. To mitigate this, we applied standard filtering techniques on the original dataset (Table \ref{tab:filtering}). Moreover, there is a possibility of the mapping algorithm retrieving erroneous matches, where the HF model name may differ from the actual one. To mitigate this, we set a higher threshold similarity for the Levenshtein function.
Threats to \textit{external validity} are related to the generalizability of our approach considering \textit{(i)} additional papers; and \textit{(ii)} other model repositories beyond \HGF. Concerning the first aspect, we expanded the initial taxonomy by querying Scopus %the Scopus digital library 
to exclusively collect peer-reviewed papers. By the second aspect, our proposed approach is adaptable to any repository that exposes a curated list of tags and textual documentation for each stored model.

 \section{Related Work}
 \label{sec:related}
Several approaches focus on predict GitHub topics.
%<<<<<<< HEAD
%Di Sipio \etal \cite{di_sipio_multinomial_2020} proposed a multi-label stochastic classifier that predicts featured topics given a README file. The same authors extended their approach by employing a collaborative filtering engine to increase the topic coverage \cite{di_rocco_hybridrec_2023}. SED-KG \cite{izadi_semantically-enhanced_2023} combines the BERT model with a topic model based on semantic relationships and knowledge graphs (KG) to enhance state-of-the-art models. GitRanking \cite{sas_gitranking_2023} exploits a semi-automated technique to create a ranked taxonomy of GitHub projects in terms of SE technologies. ZestXML \cite{widyasari_topic_2023} is an approach based on extreme multi-labeling learning conceived to outperform traditional approaches. 
%=======
Di Rocco \etal \cite{di_rocco_hybridrec_2023} conceived a hybrid recommender system that employs a multi-label stochastic classifier and collaborative filtering algorithm to predict featured topics given a README file. SED-KG \cite{izadi_semantically-enhanced_2023} combines the BERT model with a topic model based on semantic relationships and knowledge graphs (KG) to enhance state-of-the-art models. Similarly, GHTRec \cite{9590294} employs BERT to predict an initial set of topics that have been used to retrieve most similar repositories given the input one. GitRanking \cite{sas_gitranking_2023} exploits a semi-automated technique to create a ranked taxonomy of GitHub projects in terms of SE technologies. 
ZestXML \cite{widyasari_topic_2023} is an approach based on extreme multi-labeling learning conceived to outperform traditional approaches. 
%>>>>>>> de2c9e942609846e5fe61042c5b3585eb226e24a

While several studies highlight the limitation of existing PTM repositories \cite{jiang_empirical_2023,gong_what_2023,montes_discrepancies_2022}, to the best of our knowledge there has been no prior approach to automatically classify PTMs. In this context, we can redefine the challenge of categorizing generic open-source software within the framework of PTM repositories, such as \HGF. Additionally, we introduce an initial semi-automated approach to establish a mapping between generic categories and SE-specific tasks.

 %First, the authors build a knowledge graph on top of 2,234 semantic relationships by considering 863 curated topics. Afterwards, the topics are manually annotated by considering \textit{i)} the popularity and \textit{ii)} their degree in the produced graph. SED-KG is eventually employed to enhance the performances of existing recommender systems for GitHub topics. The evaluation shows that using the proposed approach leads to better performances in terms of the considered metrics. Compared to SED-KG, our approach makes use of a semi-automated technique to create a ranked  taxonomy of GitHub topics; furthermore, in their classification with find terms that are programming languages, and technologies.

\section{Conclusion and Future work}
\label{sec:Conclusion}
Aiming to map generic PTM categories with specific SE tasks, this paper envisions a semi-automated strategy based on a similarity-based algorithm. We first collected %needed 
metadata from \HGF to foster automatic categorization of the stored PTMs. Then, we built an SE taxonomy by grouping tasks that exploit PTMs into macro and sub-tasks. The mined data has been used to \textit{(i)} automatically categorize PTMs by exploited data available in the corresponding model cards and \textit{(ii)} map generic PTMs categories to corresponding SE tasks. Our findings demonstrate that our %proposed 
methodology can help developers browse \HGF even though further experiments are needed to confirm our initial results. 

For future work, we aim to improve the mapping algorithm by using well-known strategies to build software taxonomy \eg active sampling algorithm. Furthermore, we can apply our approach to additional public repositories of PTMs. Last but not least, we can envision a set of recommender systems that rely on our approach to assist developers in deploying the chosen PTMs, thus fostering the creation of an AI-based multi-agent framework tailored for SE tasks.

\section*{Acknowledgement}
%We acknowledge the following national projects: EMELIOT, PRIN 2020 program (Contract 2020W3A5FY), project PRIN 2022 PNRR FRINGE grant n. P2022553SL ``PRIN 2022'' and project TRex-SE, grant n. 2022LKJWHC.
%<<<<<<< HEAD
This work has been partially supported by the EMELIOT national research project, which has been funded by the MUR under the PRIN 2020 program grant n. 2020W3A5FY. 
The work has been also partially supported by the European Union--NextGenerationEU through the Italian Ministry of University and Research, Projects PRIN 2022 PNRR \emph{``FRINGE: context-aware FaiRness engineerING in complex software systEms''} grant n. P2022553SL. We acknowledge the Italian ``PRIN 2022'' project TRex-SE: \emph{``Trustworthy Recommenders for Software Engineers,''} grant n. 2022LKJWHC.
%=======
%This work has been partially supported by the Italian PRIN 2020 EMELIOT project (n. 2020W3A5FY), by the European Union--NextGenerationEU through the Italian Ministry of University and Research, PRIN 2022 PNRR FRINGE project (n. P2022553SL), and by the Italian PRIN 2022 TRex-SE project (n. 2022LKJWHC).
%>>>>>>> fe9b075605b80744d46d1ad0d45629bb8f177ddd

\bibliographystyle{ACM-Reference-Format}
\bibliography{bib}


\begin{thebibliography}{29}

%%% the Web address.

\ifx \showCODEN    \undefined \def \showCODEN     #1{\unskip}     \fi
\ifx \showDOI      \undefined \def \showDOI       #1{#1}\fi
\ifx \showISBNx    \undefined \def \showISBNx     #1{\unskip}     \fi
\ifx \showISBNxiii \undefined \def \showISBNxiii  #1{\unskip}     \fi
\ifx \showISSN     \undefined \def \showISSN      #1{\unskip}     \fi
\ifx \showLCCN     \undefined \def \showLCCN      #1{\unskip}     \fi
\ifx \shownote     \undefined \def \shownote      #1{#1}          \fi
\ifx \showarticletitle \undefined \def \showarticletitle #1{#1}   \fi
\ifx \showURL      \undefined \def \showURL       {\relax}        \fi
% invisible to TeX
\providecommand\bibfield[2]{#2}
\providecommand\bibinfo[2]{#2}
\providecommand\natexlab[1]{#1}
\providecommand\showeprint[2][]{arXiv:#2}

\bibitem[Ait et~al\mbox{.}(2023)]%
        {ait_hfcommunity_2023}
\bibfield{author}{\bibinfo{person}{Adem Ait}, \bibinfo{person}{Javier
  Luis~C\'{a}novas Izquierdo}, {and} \bibinfo{person}{Jordi Cabot}.}
  \bibinfo{year}{2023}\natexlab{}.
\newblock \showarticletitle{{HFCommunity}: {A} {Tool} to {Analyze} the
  {Hugging} {Face} {Hub} {Community}}. In \bibinfo{booktitle}{\emph{Procs. of
  SANER 2023}}. \bibinfo{pages}{728--732}.
\newblock
\urldef\tempurl%
\url{https://doi.org/10.1109/SANER56733.2023.00080}
\showDOI{\tempurl}
\newblock
\shownote{ISSN: 2640-7574}.


\bibitem[Buckland and Gey(1994)]%
        {buckland_relationship_1994}
\bibfield{author}{\bibinfo{person}{Michael Buckland} {and}
  \bibinfo{person}{Fredric Gey}.} \bibinfo{year}{1994}\natexlab{}.
\newblock \showarticletitle{The relationship between recall and precision}.
\newblock \bibinfo{journal}{\emph{Journal of the American society for
  information science}} \bibinfo{volume}{45}, \bibinfo{number}{1}
  (\bibinfo{year}{1994}), \bibinfo{pages}{12--19}.
\newblock
\newblock
\shownote{Publisher: Wiley Online Library}.


\bibitem[Casta\~{n}o et~al\mbox{.}(2023)]%
        {castano_analyzing_2023}
\bibfield{author}{\bibinfo{person}{Joel Casta\~{n}o}, \bibinfo{person}{Silverio
  Mart\'{\i}nez-Fern\'{a}ndez}, \bibinfo{person}{Xavier Franch}, {and}
  \bibinfo{person}{Justus Bogner}.} \bibinfo{year}{2023}\natexlab{}.
\newblock \bibinfo{title}{Analyzing the {Evolution} and {Maintenance} of {ML}
  {Models} on {Hugging} {Face}}.
\newblock
\newblock
\urldef\tempurl%
\url{https://doi.org/10.48550/arXiv.2311.13380}
\showDOI{\tempurl}
\newblock
\shownote{arXiv:2311.13380 [cs]}.


\bibitem[Chang and Lin(2011)]%
        {chang_libsvm_2011}
\bibfield{author}{\bibinfo{person}{Chih-Chung Chang} {and}
  \bibinfo{person}{Chih-Jen Lin}.} \bibinfo{year}{2011}\natexlab{}.
\newblock \showarticletitle{{LIBSVM}: {A} library for support vector machines}.
\newblock \bibinfo{journal}{\emph{ACM Trans. on Intelligent Systems and
  Technology}} \bibinfo{volume}{2}, \bibinfo{number}{3} (\bibinfo{date}{April}
  \bibinfo{year}{2011}), \bibinfo{pages}{1--27}.
\newblock
\showISSN{2157-6904, 2157-6912}
\urldef\tempurl%
\url{https://doi.org/10.1145/1961189.1961199}
\showDOI{\tempurl}


\bibitem[Di~Rocco et~al\mbox{.}(2021)]%
        {di_rocco_development_2021}
\bibfield{author}{\bibinfo{person}{Juri Di~Rocco}, \bibinfo{person}{Davide
  Di~Ruscio}, \bibinfo{person}{Claudio Di~Sipio}, \bibinfo{person}{Phuong~T.
  Nguyen}, {and} \bibinfo{person}{Riccardo Rubei}.}
  \bibinfo{year}{2021}\natexlab{}.
\newblock \showarticletitle{Development of recommendation systems for software
  engineering: the {CROSSMINER} experience}.
\newblock \bibinfo{journal}{\emph{Empirical Software Engineering}}
  \bibinfo{volume}{26}, \bibinfo{number}{4} (\bibinfo{date}{July}
  \bibinfo{year}{2021}), \bibinfo{pages}{69}.
\newblock
\showISSN{1382-3256, 1573-7616}
\urldef\tempurl%
\url{https://doi.org/10.1007/s10664-021-09963-7}
\showDOI{\tempurl}


\bibitem[Di~Rocco et~al\mbox{.}(2023)]%
        {di_rocco_hybridrec_2023}
\bibfield{author}{\bibinfo{person}{Juri Di~Rocco}, \bibinfo{person}{Davide
  Di~Ruscio}, \bibinfo{person}{Claudio Di~Sipio}, \bibinfo{person}{Phuong~T.
  Nguyen}, {and} \bibinfo{person}{Riccardo Rubei}.}
  \bibinfo{year}{2023}\natexlab{}.
\newblock \showarticletitle{{HybridRec}: {A} recommender system for tagging
  {GitHub} repositories}.
\newblock \bibinfo{journal}{\emph{Applied Intelligence}} \bibinfo{volume}{53},
  \bibinfo{number}{8} (\bibinfo{date}{April} \bibinfo{year}{2023}),
  \bibinfo{pages}{9708--9730}.
\newblock
\showISSN{0924-669X, 1573-7497}
\urldef\tempurl%
\url{https://doi.org/10.1007/s10489-022-03864-y}
\showDOI{\tempurl}


\bibitem[Di~Sipio et~al\mbox{.}(2024)]%
        {replicationPackage}
\bibfield{author}{\bibinfo{person}{Claudio Di~Sipio}, \bibinfo{person}{Riccardo
  Rubei}, \bibinfo{person}{Juri {Di Rocco}}, \bibinfo{person}{Davide {Di
  Ruscio}}, {and} \bibinfo{person}{Phuong~T. Nguyen}.}
  \bibinfo{year}{2024}\natexlab{}.
\newblock \bibinfo{booktitle}{\emph{{Replication Package: Automated
  categorization of pre-trained models for software engineering: A case study
  with a Hugging Face dataset}}}.
\newblock
\urldef\tempurl%
\url{https://github.com/MDEGroup/EASE2024-HF-ReplicationPackage}
\showURL{%
\tempurl}


\bibitem[Di~Sipio et~al\mbox{.}(2020)]%
        {di_sipio_multinomial_2020}
\bibfield{author}{\bibinfo{person}{Claudio Di~Sipio}, \bibinfo{person}{Riccardo
  Rubei}, \bibinfo{person}{Davide Di~Ruscio}, {and} \bibinfo{person}{Phuong~T.
  Nguyen}.} \bibinfo{year}{2020}\natexlab{}.
\newblock \showarticletitle{A {Multinomial} {Na\"{\i}ve} {Bayesian} ({MNB})
  {Network} to {Automatically} {Recommend} {Topics} for {GitHub}
  {Repositories}}. In \bibinfo{booktitle}{\emph{Procs. of the {Evaluation} and
  {Assessment} in {Software} {Engineering}}}. \bibinfo{publisher}{ACM},
  \bibinfo{address}{Trondheim Norway}, \bibinfo{pages}{71--80}.
\newblock
\showISBNx{978-1-4503-7731-7}
\urldef\tempurl%
\url{https://doi.org/10.1145/3383219.3383227}
\showDOI{\tempurl}


\bibitem[Dilhara et~al\mbox{.}(2021)]%
        {10.1145/3453478}
\bibfield{author}{\bibinfo{person}{Malinda Dilhara}, \bibinfo{person}{Ameya
  Ketkar}, {and} \bibinfo{person}{Danny Dig}.} \bibinfo{year}{2021}\natexlab{}.
\newblock \showarticletitle{Understanding Software-2.0: A Study of Machine
  Learning Library Usage and Evolution}.
\newblock \bibinfo{journal}{\emph{ACM Trans. Softw. Eng. Methodol.}}
  \bibinfo{volume}{30}, \bibinfo{number}{4}, Article \bibinfo{articleno}{55}
  (\bibinfo{date}{jul} \bibinfo{year}{2021}), \bibinfo{numpages}{42}~pages.
\newblock
\showISSN{1049-331X}
\urldef\tempurl%
\url{https://doi.org/10.1145/3453478}
\showDOI{\tempurl}


\bibitem[Ding et~al\mbox{.}(2022)]%
        {ding_can_2022}
\bibfield{author}{\bibinfo{person}{Zishuo Ding}, \bibinfo{person}{Heng Li},
  \bibinfo{person}{Weiyi Shang}, {and} \bibinfo{person}{Tse-Hsun~Peter Chen}.}
  \bibinfo{year}{2022}\natexlab{}.
\newblock \showarticletitle{Can pre-trained code embeddings improve model
  performance? {Revisiting} the use of code embeddings in software engineering
  tasks}.
\newblock \bibinfo{journal}{\emph{Empirical Software Engineering}}
  \bibinfo{volume}{27}, \bibinfo{number}{3} (\bibinfo{date}{March}
  \bibinfo{year}{2022}), \bibinfo{pages}{63}.
\newblock
\showISSN{1573-7616}
\urldef\tempurl%
\url{https://doi.org/10.1007/s10664-022-10118-5}
\showDOI{\tempurl}


\bibitem[Dong et~al\mbox{.}(2023)]%
        {dong2023selfcollaboration}
\bibfield{author}{\bibinfo{person}{Yihong Dong}, \bibinfo{person}{Xue Jiang},
  \bibinfo{person}{Zhi Jin}, {and} \bibinfo{person}{Ge Li}.}
  \bibinfo{year}{2023}\natexlab{}.
\newblock \bibinfo{title}{Self-collaboration Code Generation via ChatGPT}.
\newblock
\newblock
\showeprint[arxiv]{2304.07590}~[cs.SE]


\bibitem[Gong et~al\mbox{.}(2023)]%
        {gong_what_2023}
\bibfield{author}{\bibinfo{person}{Lina Gong}, \bibinfo{person}{Jingxuan
  Zhang}, \bibinfo{person}{Mingqiang Wei}, \bibinfo{person}{Haoxiang Zhang},
  {and} \bibinfo{person}{Zhiqiu Huang}.} \bibinfo{year}{2023}\natexlab{}.
\newblock \showarticletitle{What {Is} the {Intended} {Usage} {Context} of
  {This} {Model}? {An} {Exploratory} {Study} of {Pre}-{Trained} {Models} on
  {Various} {Model} {Repositories}}.
\newblock \bibinfo{journal}{\emph{ACM Trans. on Software Engineering and
  Methodology}} \bibinfo{volume}{32}, \bibinfo{number}{3} (\bibinfo{date}{May}
  \bibinfo{year}{2023}), \bibinfo{pages}{69:1--69:57}.
\newblock
\showISSN{1049-331X}
\urldef\tempurl%
\url{https://doi.org/10.1145/3569934}
\showDOI{\tempurl}


\bibitem[Han et~al\mbox{.}(2021)]%
        {HAN2021225}
\bibfield{author}{\bibinfo{person}{Xu Han}, \bibinfo{person}{Zhengyan Zhang},
  \bibinfo{person}{Ning Ding}, \bibinfo{person}{Yuxian Gu},
  \bibinfo{person}{Xiao Liu}, {et~al\mbox{.}}} \bibinfo{year}{2021}\natexlab{}.
\newblock \showarticletitle{Pre-trained models: Past, present and future}.
\newblock \bibinfo{journal}{\emph{AI Open}}  \bibinfo{volume}{2}
  (\bibinfo{year}{2021}), \bibinfo{pages}{225--250}.
\newblock
\showISSN{2666-6510}
\urldef\tempurl%
\url{https://doi.org/10.1016/j.aiopen.2021.08.002}
\showDOI{\tempurl}


\bibitem[Hong et~al\mbox{.}(2023)]%
        {hong2023metagpt}
\bibfield{author}{\bibinfo{person}{Sirui Hong}, \bibinfo{person}{Mingchen
  Zhuge}, \bibinfo{person}{Jonathan Chen}, \bibinfo{person}{Xiawu Zheng},
  \bibinfo{person}{Yuheng Cheng}, {et~al\mbox{.}}}
  \bibinfo{year}{2023}\natexlab{}.
\newblock \bibinfo{title}{MetaGPT: Meta Programming for A Multi-Agent
  Collaborative Framework}.
\newblock
\newblock
\showeprint[arxiv]{2308.00352}~[cs.AI]


\bibitem[Hou et~al\mbox{.}(2023)]%
        {hou_large_2023}
\bibfield{author}{\bibinfo{person}{Xinyi Hou}, \bibinfo{person}{Yanjie Zhao},
  \bibinfo{person}{Yue Liu}, \bibinfo{person}{Zhou Yang},
  \bibinfo{person}{Kailong Wang}, {et~al\mbox{.}}}
  \bibinfo{year}{2023}\natexlab{}.
\newblock \bibinfo{title}{Large {Language} {Models} for {Software}
  {Engineering}: {A} {Systematic} {Literature} {Review}}.
\newblock
\newblock
\urldef\tempurl%
\url{https://doi.org/10.48550/arXiv.2308.10620}
\showDOI{\tempurl}
\newblock
\shownote{arXiv:2308.10620 [cs]}.


\bibitem[Izadi et~al\mbox{.}(2023)]%
        {izadi_semantically-enhanced_2023}
\bibfield{author}{\bibinfo{person}{Maliheh Izadi}, \bibinfo{person}{Mahtab
  Nejati}, {and} \bibinfo{person}{Abbas Heydarnoori}.}
  \bibinfo{year}{2023}\natexlab{}.
\newblock \showarticletitle{Semantically-enhanced topic recommendation systems
  for software projects}.
\newblock \bibinfo{journal}{\emph{Empirical Software Engineering}}
  \bibinfo{volume}{28}, \bibinfo{number}{2} (\bibinfo{date}{Feb.}
  \bibinfo{year}{2023}), \bibinfo{pages}{50}.
\newblock
\showISSN{1573-7616}
\urldef\tempurl%
\url{https://doi.org/10.1007/s10664-022-10272-w}
\showDOI{\tempurl}


\bibitem[Jiang et~al\mbox{.}(2023)]%
        {jiang_empirical_2023}
\bibfield{author}{\bibinfo{person}{Wenxin Jiang}, \bibinfo{person}{Nicholas
  Synovic}, \bibinfo{person}{Matt Hyatt}, \bibinfo{person}{Taylor~R.
  Schorlemmer}, \bibinfo{person}{Rohan Sethi}, {et~al\mbox{.}}}
  \bibinfo{year}{2023}\natexlab{}.
\newblock \showarticletitle{An {Empirical} {Study} of {Pre}-{Trained} {Model}
  {Reuse} in the {Hugging} {Face} {Deep} {Learning} {Model} {Registry}}. In
  \bibinfo{booktitle}{\emph{Procs. of ICSE 2023}}. \bibinfo{publisher}{IEEE
  Press}, \bibinfo{address}{Melbourne, Victoria, Australia},
  \bibinfo{pages}{2463--2475}.
\newblock
\showISBNx{978-1-66545-701-9}
\urldef\tempurl%
\url{https://doi.org/10.1109/ICSE48619.2023.00206}
\showDOI{\tempurl}


\bibitem[Liu et~al\mbox{.}(2019)]%
        {liu2019roberta}
\bibfield{author}{\bibinfo{person}{Yinhan Liu}, \bibinfo{person}{Myle Ott},
  \bibinfo{person}{Naman Goyal}, \bibinfo{person}{Jingfei Du},
  \bibinfo{person}{Mandar Joshi}, {et~al\mbox{.}}}
  \bibinfo{year}{2019}\natexlab{}.
\newblock \bibinfo{title}{RoBERTa: A Robustly Optimized BERT Pretraining
  Approach}.
\newblock
\newblock
\showeprint[arxiv]{1907.11692}~[cs.CL]


\bibitem[Mitchell et~al\mbox{.}(2019)]%
        {10.1145/3287560.3287596}
\bibfield{author}{\bibinfo{person}{Margaret Mitchell}, \bibinfo{person}{Simone
  Wu}, \bibinfo{person}{Andrew Zaldivar}, \bibinfo{person}{Parker Barnes},
  \bibinfo{person}{Lucy Vasserman}, {et~al\mbox{.}}}
  \bibinfo{year}{2019}\natexlab{}.
\newblock \showarticletitle{Model Cards for Model Reporting}. In
  \bibinfo{booktitle}{\emph{Procs. of the Conf. on Fairness, Accountability,
  and Transparency}} (Atlanta, GA, USA)
  \emph{(\bibinfo{series}{FAT\textasteriskcentered '19})}.
  \bibinfo{publisher}{ACM}, \bibinfo{pages}{220–229}.
\newblock
\showISBNx{9781450361255}
\urldef\tempurl%
\url{https://doi.org/10.1145/3287560.3287596}
\showDOI{\tempurl}


\bibitem[Montes et~al\mbox{.}(2022)]%
        {montes_discrepancies_2022}
\bibfield{author}{\bibinfo{person}{Diego Montes}, \bibinfo{person}{Pongpatapee
  Peerapatanapokin}, \bibinfo{person}{Jeff Schultz}, \bibinfo{person}{Chengjun
  Guo}, \bibinfo{person}{Wenxin Jiang}, {et~al\mbox{.}}}
  \bibinfo{year}{2022}\natexlab{}.
\newblock \showarticletitle{Discrepancies among pre-trained deep neural
  networks: a new threat to model zoo reliability}. In
  \bibinfo{booktitle}{\emph{Procs. of ESEC/FSE 2022}}.
  \bibinfo{publisher}{ACM}, \bibinfo{pages}{1605--1609}.
\newblock
\showISBNx{978-1-4503-9413-0}
\urldef\tempurl%
\url{https://doi.org/10.1145/3540250.3560881}
\showDOI{\tempurl}


\bibitem[Navarro(2001)]%
        {levenshtein}
\bibfield{author}{\bibinfo{person}{Gonzalo Navarro}.}
  \bibinfo{year}{2001}\natexlab{}.
\newblock \showarticletitle{A guided tour to approximate string matching}.
\newblock \bibinfo{journal}{\emph{Comput. Surveys}} \bibinfo{volume}{33},
  \bibinfo{number}{1} (\bibinfo{year}{2001}), \bibinfo{pages}{31--88}.
\newblock
\urldef\tempurl%
\url{https://doi.org/10.1145/375360.375365}
\showDOI{\tempurl}


\bibitem[Refaeilzadeh et~al\mbox{.}(2009)]%
        {Refaeilzadeh2009}
\bibfield{author}{\bibinfo{person}{Payam Refaeilzadeh}, \bibinfo{person}{Lei
  Tang}, {and} \bibinfo{person}{Huan Liu}.} \bibinfo{year}{2009}\natexlab{}.
\newblock \bibinfo{booktitle}{\emph{Cross-Validation}}.
\newblock \bibinfo{publisher}{Springer US}, \bibinfo{address}{Boston, MA},
  \bibinfo{pages}{532--538}.
\newblock
\showISBNx{978-0-387-39940-9}
\urldef\tempurl%
\url{https://doi.org/10.1007/978-0-387-39940-9\_565}
\showDOI{\tempurl}


\bibitem[Rennie et~al\mbox{.}(2003)]%
        {rennie_tackling_nodate}
\bibfield{author}{\bibinfo{person}{Jason D~M Rennie}, \bibinfo{person}{Lawrence
  Shih}, \bibinfo{person}{Jaime Teevan}, {and} \bibinfo{person}{David~R
  Karger}.} \bibinfo{year}{2003}\natexlab{}.
\newblock \showarticletitle{Tackling the {Poor} {Assumptions} of {Naive}
  {Bayes} {Text} {Classifiers}}.
\newblock  (\bibinfo{year}{2003}).
\newblock


\bibitem[Robillard et~al\mbox{.}(2014)]%
        {robillard_recommendation_2014}
\bibfield{editor}{\bibinfo{person}{Martin~P. Robillard}, \bibinfo{person}{Walid
  Maalej}, \bibinfo{person}{Robert~J. Walker}, {and} \bibinfo{person}{Thomas
  Zimmermann}} (Eds.). \bibinfo{year}{2014}\natexlab{}.
\newblock \bibinfo{booktitle}{\emph{Recommendation {Systems} in {Software}
  {Engineering}}}.
\newblock \bibinfo{publisher}{Springer Berlin Heidelberg},
  \bibinfo{address}{Berlin, Heidelberg}.
\newblock
\showISBNx{978-3-642-45134-8 978-3-642-45135-5}
\urldef\tempurl%
\url{https://doi.org/10.1007/978-3-642-45135-5}
\showDOI{\tempurl}


\bibitem[Sas et~al\mbox{.}(2023)]%
        {sas_gitranking_2023}
\bibfield{author}{\bibinfo{person}{Cezar Sas}, \bibinfo{person}{Andrea
  Capiluppi}, \bibinfo{person}{Claudio Di~Sipio}, \bibinfo{person}{Juri
  Di~Rocco}, {and} \bibinfo{person}{Davide Di~Ruscio}.}
  \bibinfo{year}{2023}\natexlab{}.
\newblock \showarticletitle{{GitRanking}: {A} ranking of {GitHub} topics for
  software classification using active sampling}.
\newblock \bibinfo{journal}{\emph{Software: Practice and Experience}}
  \bibinfo{volume}{53}, \bibinfo{number}{10} (\bibinfo{date}{Oct.}
  \bibinfo{year}{2023}), \bibinfo{pages}{1982--2006}.
\newblock
\showISSN{0038-0644, 1097-024X}
\urldef\tempurl%
\url{https://doi.org/10.1002/spe.3238}
\showDOI{\tempurl}


\bibitem[Tufano et~al\mbox{.}(2022)]%
        {tufano_using_2022}
\bibfield{author}{\bibinfo{person}{Rosalia Tufano}, \bibinfo{person}{Simone
  Masiero}, \bibinfo{person}{Antonio Mastropaolo}, \bibinfo{person}{Luca
  Pascarella}, \bibinfo{person}{Denys Poshyvanyk}, {et~al\mbox{.}}}
  \bibinfo{year}{2022}\natexlab{}.
\newblock \showarticletitle{Using pre-trained models to boost code review
  automation}. In \bibinfo{booktitle}{\emph{Procs. of the 44th {Int.} {Conf.}
  on {Software} {Engineering}}} \emph{(\bibinfo{series}{{ICSE} '22})}.
  \bibinfo{publisher}{ACM}, \bibinfo{pages}{2291--2302}.
\newblock
\showISBNx{978-1-4503-9221-1}
\urldef\tempurl%
\url{https://doi.org/10.1145/3510003.3510621}
\showDOI{\tempurl}


\bibitem[Widyasari et~al\mbox{.}(2023)]%
        {widyasari_topic_2023}
\bibfield{author}{\bibinfo{person}{Ratnadira Widyasari},
  \bibinfo{person}{Zhipeng Zhao}, \bibinfo{person}{Thanh~Le Cong},
  \bibinfo{person}{Hong Jin~Kang}, {and} \bibinfo{person}{David Lo}.}
  \bibinfo{year}{2023}\natexlab{}.
\newblock \showarticletitle{Topic {Recommendation} for {GitHub} {Repositories}:
  {How} {Far} {Can} {Extreme} {Multi}-{Label} {Learning} {Go}?}. In
  \bibinfo{booktitle}{\emph{2023 {IEEE} {Int.} {Conf.} on {Software}
  {Analysis}, {Evolution} and {Reengineering} ({SANER})}}.
  \bibinfo{publisher}{IEEE}, \bibinfo{address}{Taipa, Macao},
  \bibinfo{pages}{167--178}.
\newblock
\showISBNx{978-1-66545-278-6}
\urldef\tempurl%
\url{https://doi.org/10.1109/SANER56733.2023.00025}
\showDOI{\tempurl}


\bibitem[Zhang et~al\mbox{.}(2022)]%
        {zhang_using_2022}
\bibfield{author}{\bibinfo{person}{Jialu Zhang}, \bibinfo{person}{Todd
  Mytkowicz}, \bibinfo{person}{Mike Kaufman}, \bibinfo{person}{Ruzica Piskac},
  {and} \bibinfo{person}{Shuvendu~K. Lahiri}.} \bibinfo{year}{2022}\natexlab{}.
\newblock \showarticletitle{Using pre-trained language models to resolve
  textual and semantic merge conflicts (experience paper)}. In
  \bibinfo{booktitle}{\emph{Procs. of ISSTA 2022}}. \bibinfo{publisher}{ACM},
  \bibinfo{pages}{77--88}.
\newblock
\showISBNx{978-1-4503-9379-9}
\urldef\tempurl%
\url{https://doi.org/10.1145/3533767.3534396}
\showDOI{\tempurl}


\bibitem[Zhou et~al\mbox{.}(2021)]%
        {9590294}
\bibfield{author}{\bibinfo{person}{Yuqi Zhou}, \bibinfo{person}{Jiawei Wu},
  {and} \bibinfo{person}{Yanchun Sun}.} \bibinfo{year}{2021}\natexlab{}.
\newblock \showarticletitle{GHTRec: {A} Personalized Service to Recommend
  GitHub Trending Repositories for Developers}. In
  \bibinfo{booktitle}{\emph{{IEEE} Int. Conf. on Web Services}}.
  \bibinfo{publisher}{{IEEE}}, \bibinfo{pages}{314--323}.
\newblock
\urldef\tempurl%
\url{https://doi.org/10.1109/ICWS53863.2021.00049}
\showDOI{\tempurl}


\end{thebibliography}


\end{document}